\documentclass[journal]{IEEEtran}
\usepackage{amsmath,amsfonts}
\usepackage{algorithmic}
\usepackage{algorithm}
\usepackage{array}
\usepackage[caption=false,font=normalsize,labelfont=sf,textfont=sf]{subfig}
\usepackage{textcomp}
\usepackage{stfloats}
\usepackage{url}
\usepackage{verbatim}
\usepackage{graphicx}
\usepackage{cite}
\usepackage{balance}
\usepackage{booktabs}
\usepackage{makecell}
\usepackage{multirow}

\begin{document}
\bstctlcite{IEEEexample:BSTcontrol}

\title{High Coupling Tunable Acoustic Resonators in Monolithic Barium Titanate}

\author{{Ian Anderson,~\IEEEmembership{Student Member,~IEEE},
Agham Posadas, Alexander A. Demkov,~\IEEEmembership{Senior Member,~IEEE}, and Ruochen Lu~\IEEEmembership{Senior Member,~IEEE}}
\thanks{Ian Anderson and Ruochen Lu are with the  Department of Electrical
and Computer Engineering, The University of Texas at Austin, Austin,
TX 78712 USA. Alexander A. Demkov is with the Department of Physics, The University of Texas at Austin, Austin, Texas 78712, USA. Agham Posadas is with La Luce Cristallina, Inc., Austin, TX 78759 USA.}
\thanks{Manuscript received April 19, 2026; revised August 16, 2026.}}



\maketitle

\begin{abstract}
The growing number of wireless communication bands has driven demand for compact, low-loss, and frequency-adjustable RF filtering. Tunable acoustic resonators are well-suited to address these needs, offering a path toward reconfigurable front ends with reduced component count. In this work, we extend upon previous conference results to investigate epitaxial barium titanate (BTO) grown on silicon as a platform for tunable acoustic resonators. We demonstrate lateral excitation of symmetric Lamb (S0) modes in 120 nm X-cut BTO membranes using a multi-cell electrode architecture that simultaneously achieves high electromechanical coupling and practical impedance levels. Devices are fabricated with laterally patterned electrodes on released BTO membranes. Under applied DC bias, ferroelectric domains align, allowing electrical excitation, frequency tuning, and quality-factor enhancement of acoustic modes. The primary resonance near 700 MHz exhibits a Bode quality factor of 175, electromechanical coupling up to 25.1\text{\%}, and series and parallel resonance tunability of 2.3\text{\%} and 5.6\text{\%}, respectively. Voltage-dependent material parameters, including permittivity, stiffness, and piezoelectric coefficients, are extracted through a combination of modified Butterworth–Van Dyke modeling and finite-element simulation to explain the observed trends. These results highlight monolithic BTO on silicon as a promising material system for laterally excited, tunable acoustic resonators for reconfigurable RF applications.
\end{abstract}

\begin{IEEEkeywords}
Acoustic Resonators, Barium Titanate, Ferroelectrics, Lamb Modes, Tunable Devices
\end{IEEEkeywords}

\section{Introduction}

Mobile communications have evolved tremendously in the past few decades. With increasing technological capabilities come increasing consumer demands, including higher data rates, smaller sizes, and lower power consumption \cite{wang_5g_2014}. As the mobile industry seeks to increase the data-processing and communication capabilities of each individual device, so too must the radio frequency (RF) front-end modules keep pace and enhance their capabilities. When it comes to selecting one signal from many in a sea of wireless traffic, the acoustic filter dominates due to its smaller size, low power consumption, and frequency-selection capabilities \cite{gong_microwave_2021,hagelauer_microwave_2018}.

Acoustic filters face a slew of difficulties in modernizing to meet current market demands. High-frequency devices are difficult to design and fabricate due to size limitations \cite{barrera_50_2026,schaffer_solidly_2023,lu_recent_2025,kramer_acoustic_2025}; a larger number of bands requires more filters, approaching nearly 100 in a single device \cite{fattinger_qqorvos_2025}. In particular, switching between these bands is quite difficult because it requires a switch for each filter, meaning that scaling the filter volume comes with more than just the cost of the individual device.

To this end, a proposed solution is to use a switchable or tunable filter across multiple frequencies. Current technologies for tunable filters generally look beyond acoustics, particularly those utilizing magneto-static filters \cite{du_frequency_2024,devitt_spin-wave_2026}, electromagnetic (EM) cavity/microstrip filters \cite{sinanis_high-q_2022,ginzberg_high-power_2023}, or RF Microelectromechanical Systems (MEMS) varactor filters \cite{cheng_high-q_2011,el-tanani_high-performance_2010}. While these solutions are ideal for larger devices, EM filters are still too large for current frequency bands, and magnetostatic filters often require strong magnetic fields to achieve adequate tuning. To date, the design of tunable RF Filters remains elusive as it requires modulation of the material stiffness, a task not easy to achieve.

\begin{figure}[!t]
\centering
\includegraphics[width=3.4in]{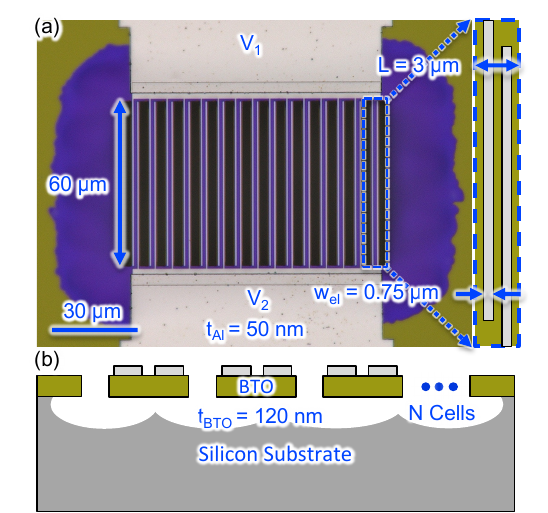}
\caption{(a) Optical image of device with 15 cells, with pictorial zoom in on single cell (b) side profile schematic highlighting layer thickness.}
\label{fig_1}
\end{figure}
For tunable acoustic resonators, an ideal candidate is the class of materials known as ferroelectrics, which are inherently piezoelectric. Ferroelectric materials differ from standard dielectrics in the nonlinearity of their dielectric permittivity. When driven by an electric field, the electrons align to screen the incoming field and saturate the polarization, resulting in a changing permittivity with applied bias and a hysteretic PE relationship. For acoustic devices, this implies changes in device coupling and stiffness through piezoelectric effects. Materials such as Scandium Aluminum Nitride (ScAlN)\cite{zou_aluminum_2022,fouladi_fbar_2019} (or ferroelectric enhanced AlBScN \cite{tong_low-field_2026}), or Lithium Niobate (LN) \cite{yang_toward_2018,gong_design_2013} are leading materials for acoustic resonators, and, what's more, they are ferroelectric; however, demonstrations of tunability in these materials remain difficult, as they require large fields or very thin films \cite{rassay_intrinsically_2021,wang_film_2020,gund_intrinsically_2022,hurtado_characterization_2025,baek_18_2025}.

Alternate materials have also been demonstrated to be suitable for tunable filters. Hafnia Zirconia (HfZrO) resonators have been demonstrated at 9 GHz using Lamb modes with relatively low applied voltage \cite{tharpe_94_2022,tharpe_instinctually_2024}. Lead Zirconium Titanate (PZT) finds applications in tunable resonators for lower frequency ranges, but fails to scale up due to high dielectric loss \cite{he_switchable_2019}. Barium Strontium Titanate (BST) has found substantial success with film bulk acoustic wave resonator (FBAR) designs in the GHz range, showing high quality factor ($Q$) and coupling, with practical tunable filter designs \cite{zolfagharloo_koohi_reconfigurable_2020,zolfagharloo_koohi_compact_2018,koohi_high_2017,nam_switchless_2021,berge_tunable_2007}. Its ferroelectric counterpart, Barium Titanate (BTO), exhibits greater capacitive tuning due to larger nonlinearities in permittivity, enabling broader tuning ranges \cite{kim_nature_2023}. Unfortunately, these high-capacitance materials (BTO, BST, PZT, etc.) generally suffer from performance degradation at high frequencies due to challenges such as low material quality, high dielectric loss, and process integration difficulties.

In recent years, it has been shown that high-quality BTO can be grown epitaxially on Silicon (Si) using a Strontium Titanate (STO) buffer layer for strain management \cite{kim_crystal_2025,posadas_thick_2021}. Given their typical use in optical and ferroelectric devices, such films are likely ideal candidates for the design of acoustic resonators \cite{kim_thermo-optic_2026,raju_highq_2025}. 

\begin{figure}[!t]
\centering
\includegraphics[width=3.3in]{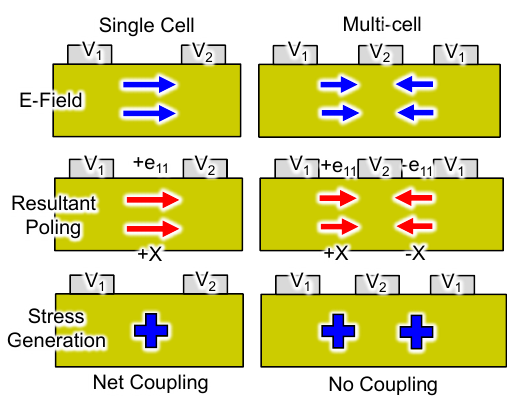}
\caption{Schematic representation of electrode design, highlighting how we have no coupling for designs with more than one cell.}
\label{fig_13}
\end{figure}

In this paper, we build on previous conference work \cite{anderson_tunable_2026} in which we demonstrated initial results for single-cell bar-type acoustic resonators, achieving a coupling of 3\text{\%} or 8\text{\%}, depending on mode, and good tunability of around 3\text{\%} or 1\text{\%}. Here, to achieve much higher-coupling devices closer to 50$\Omega$, we extend the design using the same stack, aiming for a parallel device configuration at higher frequencies. By designing electrode placement to better couple into our stress profile, we achieve over 3x the coupling of previous devices, up to 25.1\text{\%}, with series and parallel resonance tunability of 2.3\text{\%} and 5.6\text{\%}, respectively. 

The paper is organized as follows. Section \ref{sec2} describes the fabrication process, focusing on device design and stack characterization. In Section \ref{sec3}, devices are characterized across their voltage-tuning range and tested for power handling and temperature stability. Section \ref{sec4} explains these results by varying material parameters to account for shifts in resonance frequency and voltage-dependent coupling behavior. Section \ref{sec5} then uses this model to implement an alternative split-electrode design that increases the resonator body size without requiring precise etched hole arrangements. Section \ref{sec7} discusses the intrinsic loss extraction with acoustic delay lines. Section \ref{sec6} details the comparison to State-of-the-Art (SoA) tunable acoustic resonators. 


\begin{figure}[!t]
\centering
\includegraphics[width=3.15in]{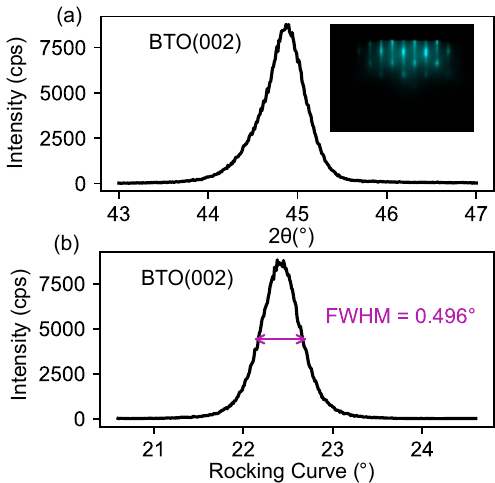}
\caption{XRD analysis figures showing (a) 2$\theta$ scans highlighting out-of-plane lattice constant and orientation, and (b) rocking curve for material quality.}
\label{fig_2}
\end{figure}

\section{Device Design and Fabrication}\label{sec2}

The device design follows the configuration shown in Figure~\ref{fig_1}, which is a 15-cell version of the previously used single-cell design \cite{anderson_tunable_2026}. This design utilizes a single pair of electrodes to couple into the fundamental mode of the 3 $\mu$m width and 60 $\mu$m aperture bar-type resonator. Electrodes are 750 nm wide and spaced 1.35 $\mu$m apart to maximize coupling, while each parallel resonator is spaced by 3 $\mu$m. The side profile in Figure~\ref{fig_1}(b) highlights our suspended cavity, our thickness of 120 nm for BTO and 50 nm for aluminum electrodes, and the cell spacing, where we use N = 15 cells to reduce impedance. In operation, a DC bias applied across the electrode pair locally poles the BTO film, enabling piezoelectric coupling of RF signals into Lamb modes through the same electrodes.

Due to the poling dynamics of BTO, the standard quarter-wavelength IDT configuration used in overmoded lateral devices does not scale well with device size. In a conventional IDT, adjacent electrodes carry opposite RF polarities, which, in a normal piezoelectric, creates an alternating stress that couples efficiently into lateral modes. However, in a ferroelectric such as BTO, the applied bias also sets the local poling direction: regions under oppositely directed fields develop equal and opposite piezoelectric coefficients $e_{ij}$, and the net stress cancels, yielding zero coupling, as illustrated in Figure~\ref{fig_13}. To counter this, the resonator body is divided into $N$ identical half-wavelength cells, each containing a single electrode pair, with etched trenches isolating adjacent cells both mechanically and electrically. Because every cell sees the same bias orientation, the piezoelectric contributions add constructively across the array, preserving high coupling into the fundamental symmetric (S0) Lamb mode near 700~MHz while reducing the device impedance to approximately 50~$\Omega$ \cite{smith_analysis_1969}.

\begin{figure}[!t]
\centering
\includegraphics[width=3.25in]{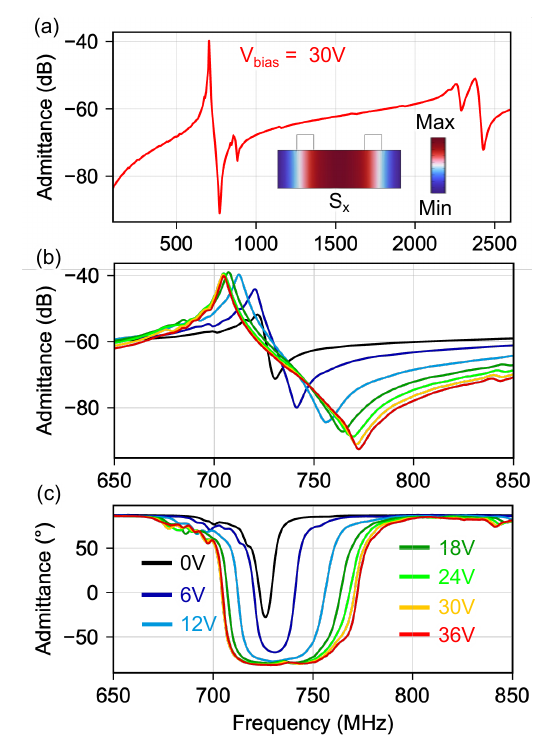}
\caption{(a) Example admittance magnitude response at V$_{bias}$ = 30V with COMSOL simulated mode profile of S0 mode (b) increasing voltage plots for magnitude and (c) phase showing progression of applied bias.}
\label{fig_3}
\end{figure}
For this study, intrinsic silicon wafers (R$\approx$ 10000 $\Omega$-cm) with 2” diameter were used as substrates. Before BTO deposition, a 5 nm-thick SrTiO$_3$ (STO) buffer layer was deposited on the clean Si surface by molecular beam epitaxy and subsequently transferred under vacuum to a sputtering system. The BTO layer was deposited by off-axis RF magnetron sputtering at a substrate temperature of 700°C and was grown to a thickness of 120 nm. Epitaxial growth was confirmed using reflection high-energy electron diffraction (RHEED) and X-ray diffraction (XRD) in Figure~\ref{fig_2}. The 120-nm BTO films on bulk Si showed an out-of-plane lattice constant of 4.036 \r{A}. Further details on growth methods and techniques can be found in \cite{posadas_thick_2021}.

The subsequent fabrication steps following growth involved etching the film using purely physical argon-based ion milling, and e-beam evaporation of the device electrodes with 50 nm of aluminum (Al). After these steps, devices are released using XeF$_2$, providing a free boundary condition for confinement of acoustic waves.

\section{Measurement Results}\label{sec3}
\subsection{Measurement and Modeling}

\begin{figure}[!t]
\centering
\includegraphics[width=3.15in]{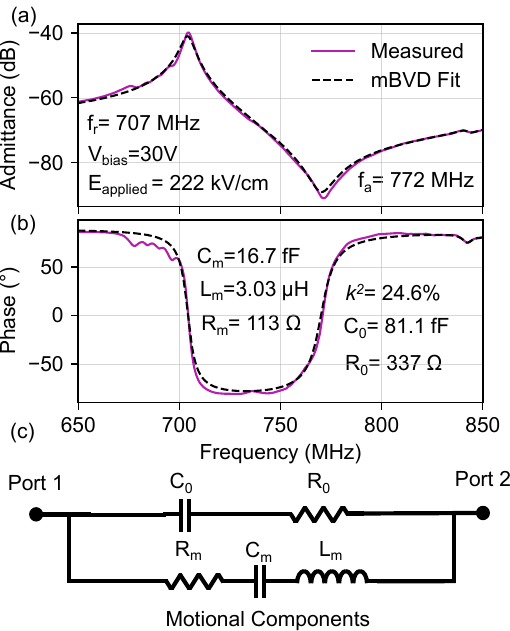}
\caption{(a) Zoomed-in admittance measurement at V$_{bias}$ = 30V with mBVD fit overlay and (b) phase, along with (c) the mBVD equivalent circuit and extracted parameters.}
\label{fig_4}
\end{figure}

Devices are measured in a two-port configuration utilizing an N5028A Vector Network Analyzer (VNA). These connections then use a DC bias tee to apply a bias between the Signal 1 and Signal 2 ports of the device. When the material is initially grown, its ferroelectric nature causes the crystal to settle into ferroelectric domains, as illustrated in similar literature \cite{ocenasek_dynamics_2023}. This means that the in-plane orientation of each domain, i.e., the c-axis of our stack, is either parallel or perpendicular to the wafer flat, depending on the orientation of the underlying silicon. With no bias, these points are in random directions, and piezoelectric coupling cancels out, giving zero (or very little) net coupling into mechanical modes. With applied bias, the domains align to produce a net polarization, as shown in Figure~\ref{fig_3}. 

Figure~\ref{fig_3}(a) shows the prime example of our acoustic mode with 30V of applied DC bias at 700 MHz, alongside a COMSOL simulated stress profile of our S0 mode, showing how we align our electric field with the stress field in our device. Due to the nature of our device, which uses a single electrode, coupling into higher-order modes is quite easy, though they inherently have lower quality factors and coupling coefficients. Figure~\ref{fig_3}(b) and (c) show the magnitude and phase of the device with increasing applied voltage, showing a few key changes in the device performance. Firstly and most expected, the device capacitance changes dramatically. This behavior arises from polarization saturation in BTO, making the relationship between the electric field and polarization highly nonlinear. Also notable trends are a decrease in the series resonance frequency, an increase in the parallel resonance and electromechanical coupling ($k^2$), and an increase in the quality factor, which will be discussed further below.

To extract device parameters, we use the modified Butterworth Van-Dyke (mBVD) model to emulate our measurements from circuit parameters. Figure~\ref{fig_4} shows an example of our model applied to the data along with the circuit parameters. Figure~\ref{fig_4}(c) shows the actual circuit we use, showing the motional components in the bottom branch consisting of resonant L$_m$ and C$_m$ giving rise to series resonance and R$_m$ modeling our motional loss. In the top branch, we have our electrical components, with C$_0$ and R$_0$ representing the dielectric capacitance and loss. The following model shows a resonance frequency of 707 MHz and an anti-resonance frequency of 772 MHz, and a modeled coupling of 24.6\text{\%}. When applied bias is -39V, a high of 25.1\text{\%} is achieved.

\subsection{Device Performance Versus Bias Voltage}

\begin{figure}[!t]
\centering
\includegraphics[width=3.45in]{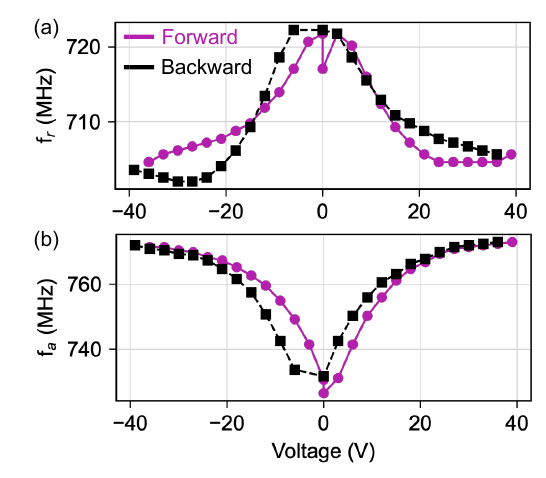}
\caption{(a) Resonance frequency variation with applied bias and (b) antiresonance frequency variation, showing hysteretic behavior.}
\label{fig_5}
\end{figure}

\begin{figure}[!t]
\centering
\includegraphics[width=3.1in]{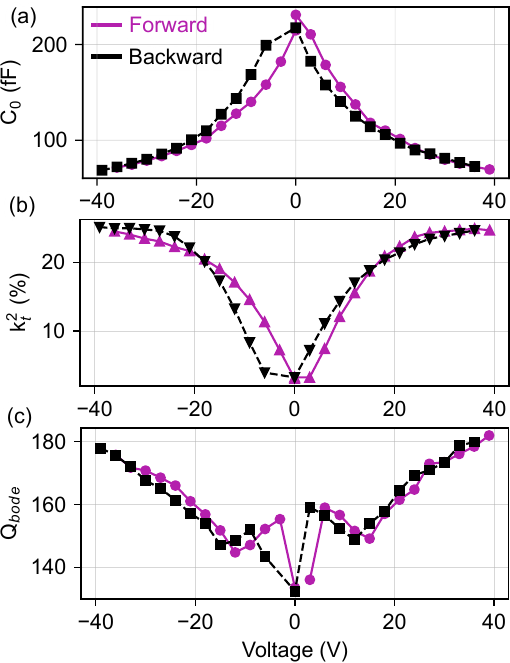}
\caption{(a) mBVD derived static capacitance and its variation with applied voltage, (b) mBVD derived coupling variation, and (c) Bode Q variation.}
\label{fig_6}
\end{figure}

Figures~\ref{fig_5} and~\ref{fig_6} show how our device performs over the range of voltages for forward and backward sweeps. Firstly, we observe that these sweeps are unequal due to the ferroelectric behavior of our material. The results of our sweeps will depend on the material's initial state or polarization. It can be seen from Fig.~\ref{fig_5}(a) that the series resonance frequency largely decreases and then slowly plateaus after around 20V, while our parallel resonance in Fig.~\ref{fig_5}(b) shows similar behavior in the opposite direction, moving a much larger amount (and notably following a different shape). Also notable is the shift in the shape around $-$3V, likely indicating our coercive field strength. Our electrode gap is 1.35 $\mu$m, meaning our coercive field value is around 2 MV/m. Further efforts will focus on measuring hysteretic loops to confirm this behavior. We can use our mBVD-derived circuit parameters to further probe our device's performance over voltage.

Figures~\ref{fig_6}(a)-(c) show the capacitance change, the coupling changes, and the quality factor ($Q$) derived from the Bode Q formula \cite{feld_after_2008}. The most important change is the change in permittivity, which is reflected in the change in capacitance. Our static capacitance value changes from 229 fF at 0V to 68.5 fF at 39V, indicating a 3x change in permittivity. We can use Equation 1 below to explain how this behavior impacts other device parameters.
\begin{equation}
    k^2 \approx \frac{e_{33}^2}{c_{33}\epsilon_{3}}
\label{eq:Coupling}
\end{equation}

Where we use the index of 3 to indicate we are using x-axis orientation, and c$_{33}$ as an approximation for our effective stiffness in the S0 mode of X-Cut. With the radical decrease in capacitance, we expect a large increase in device coupling according to the above equation, and indeed this is what we observe, as the coupling follows a shape very similar to the inverse of the permittivity. Also, following this shape is the parallel resonance frequency, as expected. For lateral field excitation (LFE) devices, the parallel resonance frequency is generally spaced from the series resonance by the amount of the electromechanical coupling \cite{rosenbaum_bulk_1988}. Though the anti-resonance frequency and coupling seem to plateau as the capacitance continues to change, indicating a change in other parameters, as will be discussed in the next section. Lastly, we can see the quality factor initially jumps to 150 and then slowly increases to 175 over the rest of the voltage range. Preliminary results on acoustic delay line structures indicate that the loss is limited by intrinsic material mechanisms, as shown in Section~\ref{sec7}.
\begin{figure}[!t]
\centering
\includegraphics[width=3.35in]{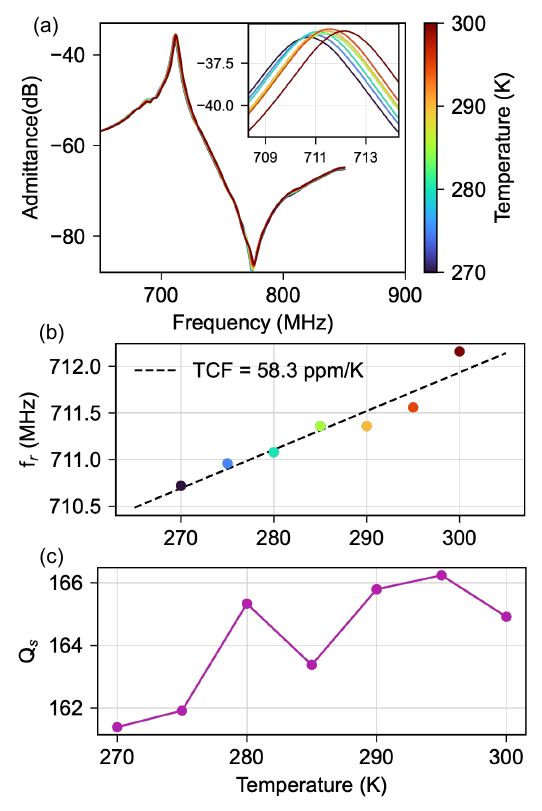}
\caption{(a) Admittance response over temperature with inset series resonance shift, (b) series resonance frequency versus temperature, and (c) series quality factor versus temperature.}
\label{fig_7}
\end{figure}

\subsection{Thermal Performance and Power Handling}
Before proceeding, it is important to establish the device's thermal performance and power handling. The following measurements are taken with devices with 5 more unit cells, hence the slightly smaller impedance values. Figure~\ref{fig_7}(a) shows our device change in admittance over a range of temperatures near and below room temp, with an inset showing a zoom-in of the series resonance. Temperatures are tested at lower ranges to avoid stress issues with very thin films, which can be mitigated in future fabrications with thicker films. We can plot the temperature versus frequency in Figure~\ref{fig_7}(b), highlighting the temperature coefficient of frequency (TCF) of 58.3 ppm/K. We can also measure the series quality factor of these modes in Figure~\ref{fig_7}(c), showing a very small trend of decreasing quality factor, but overall a consistent value near our Bode $Q$ extracted value.

The standout in our TCF data is the positive value, i.e., an increase in frequency with increasing temperature. Generally, one would expect a decrease with increasing temperature, as the body expands and the wavelength increases, as is typical of other acoustic resonators. In fact, similar literature on the TCF of BST devices reports negative TCF values \cite{koohi_high_2017}. Positive TCF values can arise from temperature-dependent stiffness parameters, but the exact reason we observe positive TCF relative to other devices in the literature is likely due to the temperature-dependent phase transitions of BTO, where the positive TCF is achieved in the tetragonal ferroelectric phase\cite{berge_field_2008}. It has been reported that BST can also exhibit this same phenomenon, but it occurs at much lower temperature, as the crystal must transition from cubic to tetragonal \cite{koohi_bst_2017,berge_field_2008}.

\begin{figure}[!t]
\centering
\includegraphics[width=3.35in]{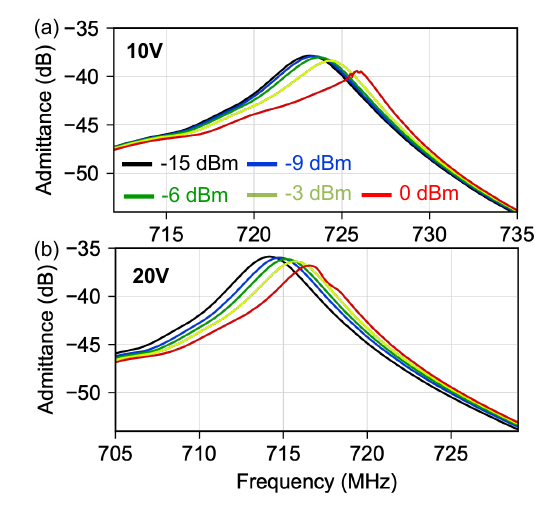}
\caption{Power handling measurements showing admittance near resonance with increasing RF input at (a) V$_{bias}$ = 10V and (b) V$_{bias}$ = 20V.}
\label{fig_8}
\end{figure}

Associated with the thermal performance is the device's power handling. Materials with low thermal conductivity (e.g., BTO, LN, or LT) exhibit lower power handling due to Joule heating because of their high thermal resistance \cite{he_heat_2004}. This is especially true in thin film released devices such as ours, which increase the magnitude of thermal nonlinearity even more. Figures~\ref{fig_8}(a) and (b) show the power handling of the device with increasing power levels and voltages of 10V and 20V. As you drive a device with higher power, we see the resonance shift toward higher frequencies, corresponding to an increase in temperature. For lower voltage, the device is slightly better at power handling due to a lower quality factor, but both devices show significant mode bending at 0 dBm. This power handling is likely to be improved with thicker films and larger device size to increase air conduction.

\section{Material Property Extraction}\label{sec4}

\begin{figure}[!t]
\centering
\includegraphics[width=3.1in]{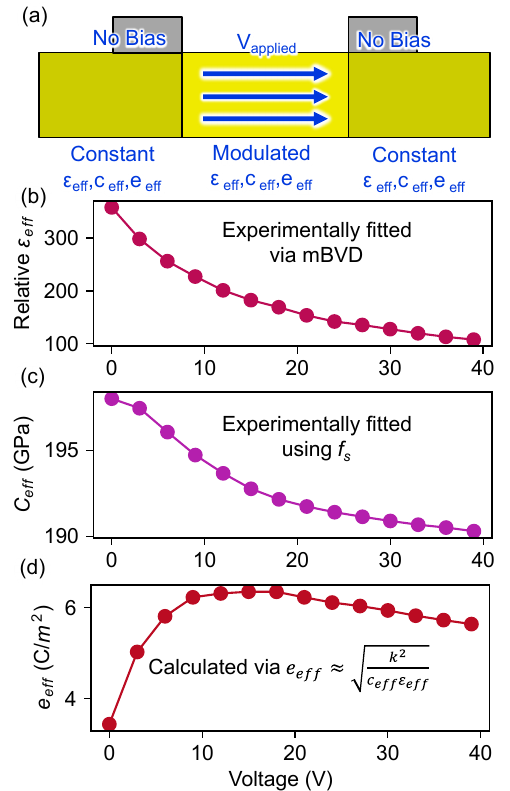}
\caption{(a) Schematic of modulated and unmodulated regions under bias, (b) relative permittivity $\epsilon_3$, (c) stiffness $C_{eff}$, and (d) piezoelectric coefficient $e_{eff}$}
\label{fig_9}
\end{figure}

Figure~\ref{fig_9} shows our extracted parameters for predicting how device behavior varies across the applied voltage range.  Figure~\ref{fig_9}(a) shows how we set up our material modulation in accordance with the applied field. We only bias the material between the two electrodes, so we only change the material parameters there, while outside this region, the material parameters are those of V$_{bias}$ = 0V. Figure~\ref{fig_9}(b) is the permittivity extracted from the capacitance values of the mBVD data. We simulate the theoretical capacitance-to-permittivity relationship in COMSOL by varying the permittivity, exporting the data to our mBVD model, and then extracting the linear relationship between capacitance and permittivity. From this, we plug in our real data and interpolate to get the permittivity.

The shift in the series resonance frequency is not attributable to the change in coupling. COMSOL simulations of changes in permittivity and piezoelectric parameters show slight changes in the series frequency through $k^2$, changing the effective stiffness $ c_{eff}$, but not only are these changes too small, but they are also in fact in the opposite direction of the measured shift. This suggests the material stiffening through other means. We model this through a change of the C$_{eff}$ parameter to decrease the resonance frequency, following the same shape as seen in Figure~\ref{fig_9}(c). It should be noted that the origin of this change in the series resonance frequency remains unknown, and is speculated to be electrostriction \cite{vendik_modeling_2008,lanz_piezoelectric_2004} due to the nature of opposing our parallel frequency shift, where instead of shifting our anti-resonance frequency in thickness field devices, it shifts our series frequency due to LFE.

\begin{figure}[!t]
\centering
\includegraphics[width=3.1in]{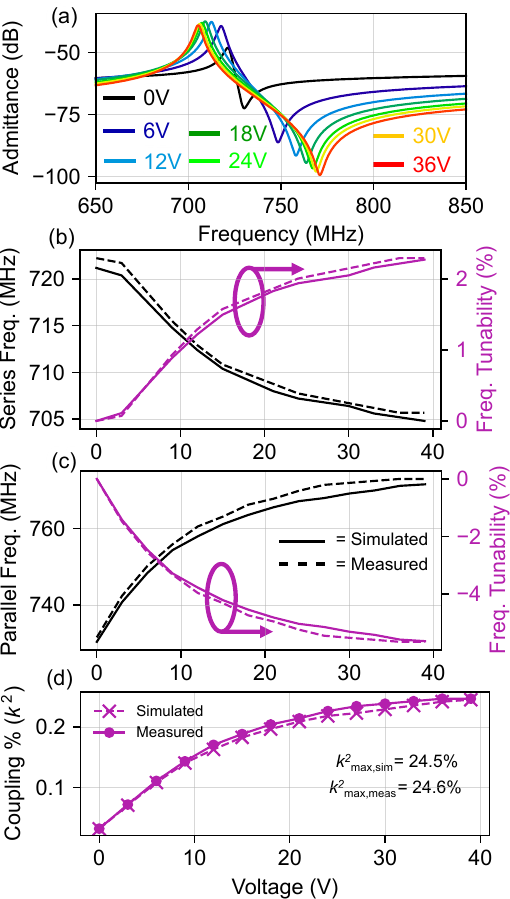}
\caption{Simulation results, showing (a) simulated admittance profiles similar to those in Fig. 5 (b) series frequency variation against measurement, along with tunability and (c) parallel resonance frequency variation with tunability (d) simulated versus measured coupling.}
\label{fig_10}
\end{figure}

Finally, in Figure~\ref{fig_9}(d), to match the parallel resonance and coupling, we add the coefficient e$_{eff}$. It can be seen that, at low voltages, the coupling is quite small and increases with domain-wall alignment. For higher voltages, it drops off to lower values. If we do not add this taper to the piezoelectric coefficient, coupling should keep rising quite fast with the change in permittivity. However, we do not observe this rise, so it is likely that an effect is causing a gradual drop in the piezoelectric coefficient.

We can see the effect this has by simulating these material parameters in COMSOL for their admittance response in Figure~\ref{fig_10}(a). Our curves follow very similarly to those in experimental results from Figure~\ref{fig_3}(b), following experimental derivation of material parameters. To better compare, we plot the series and parallel resonance frequencies in Figures~\ref{fig_10}(b) and (c), showing exactly how our frequencies change over time. Alongside these frequencies is their measured tunability, as the fractional frequency shift relative to the zero-applied-bias mode. This shows a series resonance frequency change of around 2.3\text{\%} and a parallel resonance frequency change of around 5.6\text{\%} owing to the change in coupling, with the changes likely to be even greater if not for the stiffness modulation. Figure~\ref{fig_10}(d) lastly shows how our simulated coupling and measured coupling differ, where the difference stems from the fact that we derived our coupling from simulation via the resonance frequency locations, which slightly underestimates in comparison to the mBVD model.

\begin{figure}[!t]
\centering
\includegraphics[width=3in]{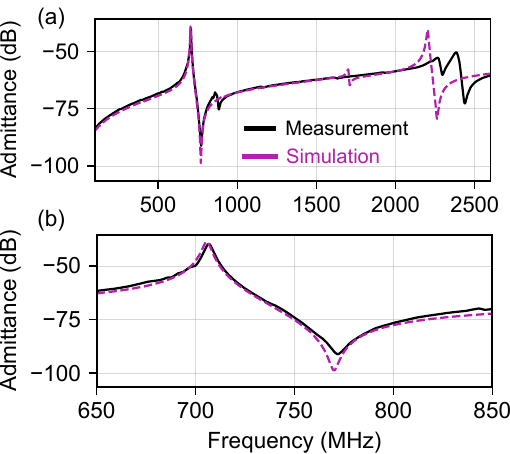}
\caption{(a) Wideband and (b) Narrowband COMSOL simulation fit to measurement data.}
\label{fig_12}
\end{figure}

Figure~\ref{fig_12}(a) and (b) show how our measurement fits to our data, showing an excellent fit for the data in the band of our simulated resonance, which is to be expected. Our third-order overtone matching is less effective, as the simulation assumes a sharp boundary between the poled and unpoled regions [Figure~\ref{fig_9}(a)], whereas in practice the fringing field likely produces a gradual transition in material properties near the electrode edges. This discrepancy is more pronounced for higher-order modes, whose shorter lateral wavelengths make them more sensitive to the spatial profile of the poling gradient. Nevertheless, the close agreement for the fundamental mode confirms that the extracted material parameters provide a reliable baseline for predicting device behavior under bias.

\section{Overmoded Acoustic Resonators}\label{sec5}

\begin{figure}[!t]
\centering
\includegraphics[width=3.3in]{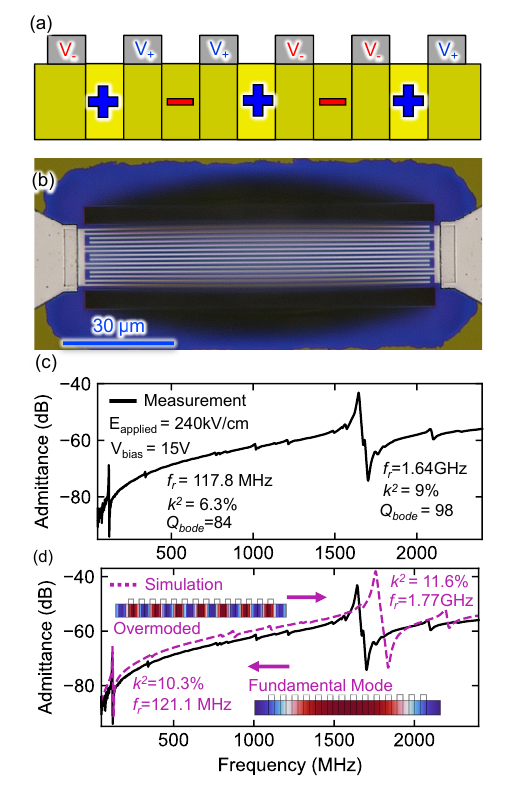}
\caption{Split-electrode overmoded resonator: (a) schematic of split-electrode stress coupling, (b) optical image of the device, (c) measured admittance at V$_{bias}$ = 15V showing fundamental and overmoded modes with extracted parameters, and (d) simulated admittance comparison.}
\label{fig_11}
\end{figure}

An alternative to the etched-trench isolation used in Section~\ref{sec2} is a split-electrode configuration, in which each signal and ground line is replaced by a pair of narrower electrodes at half the original pitch. This restores uniform poling without requiring precise trench placement, but couples into only half of the lateral stress profile, reducing $k^2$. 

Figure ~\ref{fig_11}(a) shows a schematic of this, illustrating how we couple into only one polarity of stress (the positive in this illustration), and leave the negative portions uncoupled. These electrodes are commonly used in traveling-wave configurations, as they don't satisfy the Bragg reflection conditions \cite{anderson_solidly_2025}. Figure~\ref{fig_11}(b) shows a device image with 3 cells, where the periodicity of the device is 5 $\mu$m, and each electrode has a width of 625 nm. For this reason, these devices will have double the field strength for a given applied bias. Figure~\ref{fig_11}(c) shows the results of such measurements, showing two modes at 120 MHz and 1.65 GHz. Due to the poling nature, our electrodes generate stress at half the normal size, coupling into our third-order lateral harmonic. With a larger wavelength, we can use a much higher frequency of 1.65 GHz (this, of course, comes at the cost of reduced coupling and a smaller critical dimension). We also couple into our lower frequency mode, or our structure's fundamental S0 mode, due to the nature of the applied stress profile. Figure~\ref{fig_11}(d) shows our COMSOL simulations of this mode, showing that we are off in frequency and capacitance similar to our previous higher-order mode from the original devices. Our permittivity is off by about 30\text{\%}, while our stiffness differs by about 10\text{\%}. The difference in coupling is likely due more to electrode misalignment in the lithographic process, as is typical for acoustic resonator devices.
 \begin{figure}[!t]
\centering
\includegraphics[width=3.1in]{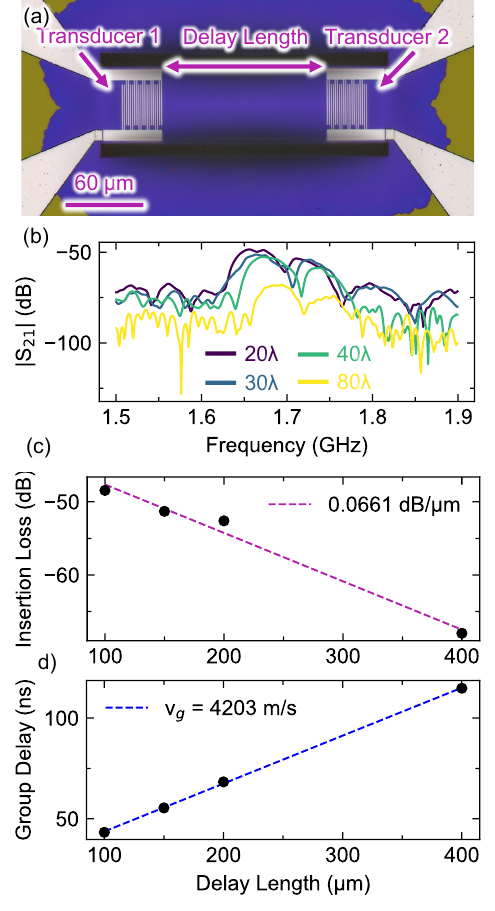}
\caption{Acoustic delay line measurements: (a) device image, (b) time-gated S$_{21}$ for varying delay lengths, (c) propagation loss extraction, and (d) group velocity extraction.}
\label{fig_adl}
\end{figure}
These discrepancies are consistent with those observed for the third-order overtone of the bar-type resonator in Section~\ref{sec4}, where the simple sharp-boundary poling model becomes less accurate at shorter wavelengths. Despite these offsets, the simulation captures the correct mode structure and relative coupling trends, indicating that the material parameters extracted in Section~\ref{sec4} provide a reasonable first-order description. More broadly, these results demonstrate that the BTO platform can support multiple resonance frequencies on a single die through lithographic control of the electrode layout alone, without requiring changes to the film or additional processing steps.

 \section{Loss Extraction with Acoustic Delay Lines}\label{sec7}

 Lastly, we present work from our acoustic delay line studies. Figure~\ref{fig_adl} shows the results of this work following a process similar to that in Ref. \cite{tsai_low_2023,chang_59_2025}, thus the explanation and extraction of material properties here will be kept brief. Device images are shown in Figure~\ref{fig_adl}(a), showing the delay line structure with two sets of transducers, one for sending and one for receiving acoustic waves. The devices used for this study have a slightly larger wavelength ($\lambda$ = 5$\mu$m), and, due to the split-electrode configuration, we couple into the third-order lateral mode, which means our frequency is much higher, around 1.65 GHz. Devices are first measured for their transmission (S$_{21}$) values. Figure~\ref{fig_adl}(b) shows the time-gated measurements of the device, highlighting the passband and the increase in loss with the increasing device delay length. We then use the gradient of the insertion loss with respect to delay length to derive the propagation loss, and the gradient of the group delay with respect to delay length to obtain the group velocity. Using these two parameters, along with the frequency, we can derive the quality factor associated with the film to be Q = 169 using the equation below:
\begin{equation}
    Q= \frac{f_0\pi}{PL_{dis}v_g}
\label{eq:loss}
\end{equation}
\begin{table}[t]
\caption{State-of-the-art ferroelectric resonators}
\label{tab:soa}
\centering
\footnotesize
\setlength{\tabcolsep}{4pt}
\renewcommand{\arraystretch}{1.12}
\begin{tabular}{@{}c c c c c c c c@{}}
\toprule
Ref. & Material & Excitation & $f$ (GHz) & $Q$ & $k^{2}$ & Tuning &
\makecell[c]{On-chip\\multi-$f$} \\
\midrule
\cite{wang_film_2020}        & ScAlN & Thickness & 2.9  & 210  & 18.1\% & 1.1\% & N \\
\cite{koohi_high_2017}       & BST   & Thickness & 2.1  & 360  & 8.6\%  & N/A   & N \\
\cite{lee_intrinsically_2013}& BTO   & Thickness & 1.65 & 160  & 2\%    & 1.8\% & Y \\
\cite{sis_intrinsically_2016}& BST   & Thickness & 0.75 & 99   & 0.1\%  & N/A   & N \\
\cite{baek_18_2025}          & LN    & Thickness & 16.3 & 98   & 11.8\% & N/A   & N \\
\cite{tharpe_94_2022}        & HfZrO & Thickness & 9.43 & 314  & 0.31\% & N/A   & N \\
\cite{anderson_tunable_2026} & BTO   & Lateral   & 0.7  & 150  & 3\%    & 1.1\% & Y \\
\midrule
\multirow{2}{*}{\makecell[c]{\textbf{This}\\\textbf{Work}}} & \textbf{BTO} & \textbf{Lateral} & \textbf{0.7}  & \textbf{175} & \textbf{25\%} & \textbf{5.6\%} & \textbf{Y} \\
 & \textbf{BTO} & \textbf{Lateral} & \textbf{1.65} & \textbf{98}  & \textbf{9\%}  & \textbf{N/A} & \textbf{Y} \\
\bottomrule
\end{tabular}
\end{table}
The quality factor can de-embed non-material-related losses, such as electrical losses, metal damping, or air damping. Because this value is so close to that of our measured resonator $Q$ it is likely the material that is limiting our device performance. As to whether this material-related loss is associated with general acoustic loss (such as Akhiezer or thermoelastic damping), or caused by our ferroelectric domain wall interactions \cite{defay_tunability_2011}, remains to be seen. It is difficult to speculate on which of these mechanisms is dominant, but future studies will focus on growing thicker films and performing frequency-dependent analysis of $Q$ to further hone in on the leading cause.

\section{Comparison to State of the Art}\label{sec6}

We compare to SoA tunable acoustic resonators in Table~\ref{tab:soa}. The most notable result is an electromechanical coupling of 25\%, the highest reported among tunable ferroelectric resonators to date. This high coupling is a direct consequence of the large permittivity tunability of BTO, as discussed in Section~\ref{sec4}. The 6\% total tuning range (combining series and parallel frequency shifts) also exceeds that of other entries in the table.

In terms of quality factor, our Bode $Q$ of 175 is comparable to those of BST and ScAlN devices, and is quite competitive given the 120~nm film thickness used in this work. As discussed in Section~\ref{sec7}, preliminary delay line measurements suggest that device $Q$ is currently limited by intrinsic material loss rather than electrode or anchor-related mechanisms, indicating clear room for improvement with thicker films.

A key distinction of this platform is the use of lateral field excitation, which enables on-chip multi-frequency operation through lithographic control of the electrode pitch. This is different from the thickness-mode devices, where the resonance frequency is set by the film thickness. Compared to LN and HfZrO, our devices operate at a lower frequency. Future iterations will address this by reducing the electrode pitch and exploring thickness-direction modes.

\section{Conclusion}\label{sec8}

This work demonstrates epitaxially grown BTO on Si as a viable platform for laterally excited tunable acoustic resonators. A multi-cell electrode architecture achieves a record electromechanical coupling of 25.1\% among tunable ferroelectric resonators while maintaining practical impedance levels. Voltage-dependent material parameters were extracted through combined mBVD modeling and finite-element simulation, providing a predictive framework for device behavior under bias. Delay line measurements confirm that loss is currently dominated by intrinsic film mechanisms. The split-electrode topology further demonstrates the platform's ability to support multiple frequencies on a single die through lithographic control alone. Future work will target thicker films for higher quality factor and power handling, alternative crystal orientations to access higher-frequency modes, and frequency-dependent loss studies to identify the dominant dissipation mechanisms.

\section*{Acknowledgments}
This work was supported by the NASA Space Technology Graduate Research Opportunity (NSTGRO), National Science Foundation (NSF) under CAREER Award No.\ 2339731, and the Office of Naval Research (Grant No. N00014-24-1–2063). The authors would also like to thank Dr. Giovanni Esteves, Dr. Tzu-Hsuan Hsu, and Ziqian Yao for helpful discussions.

\balance
\bibliographystyle{IEEEtran}
\bibliography{IEEESettings,references}

\end{document}